\documentstyle[12pt,epsfig]{article}
\def\reference{\parskip 0pt\par\noindent\hangindent 0.5 truecm}

\begin{document}
%
%
\title{High Frequency Peakers}
%


\author{Daniele Dallacasa $^{1}$ 
} 

\date{}
\maketitle

{\center
$^1$ Dip. di Astronomia, Univ. of Bologna, Via Ranzani 1,
 I-40127, Bologna, Italy \\ ddallaca@ira.cnr.it \\[3mm]
}

%
\begin{abstract}
There is quite a clear anticorrelation between the intrinsic peak
frequency and the overall radio source size in Compact Steep Spectrum
(CSS) and GHz Peaked Spectrum (GPS) radio sources as shown by O'Dea
(1998).  
This feature is interpreted in terms of synchrotron self-absorption
(although free-free absorption may play a role as well) of the
radiation emitted by a small radio source which is growing within the
inner region of the host galaxy. This leads to the hypothesis that
these objects are young and that the radio source is still
developing/expanding within the host galaxy itself. \\
Very young radio sources must have the peak in their radio spectra
occurring above a few tens of GHz, and for this reason they are termed
High Frequency Peakers (HFPs). These newly born radio sources must be
very rare given that they spend very little time in this stage. \\ 
H$_o$ = 100 km\,s$^{-1}$Mpc$^{-1}$ and q$_o$=0.5 are used throughout
this paper. 

\end{abstract}


\bigskip

%
%

\section{Introduction}

The number of radio sources that appear unresolved at arcsecond scale
resolution is rather conspicuous in all modern catalogues. The radio
spectra of these sources can be separated 
into two distinct classes: the flat spectrum objects and the steep
spectrum ones. Other properties are related to this first distinction:
flux density and polarisation variability, core dominance, which
can also be extended to other observing bands from radio frequencies
through to X-rays. 
All these characteristics can be interpreted in terms of Unified Scheme
models (e.g. Urry \& Padovani 1995) where a relativistic jet is
amplified when its direction is aligned to the line of sight. From 
this picture it is then possible to infer that a substantial fraction
of the ``small'' sources is indeed shortened by projection and, in
particular, the blazars (flat spectrum radio quasars and BL Lacs) are
well known in this sense. It is also clear from this simple picture
that intrinsically small radio sources are expected to be hosted in
galaxies i.e. where the radio source major axis is at a large
angle with respect to the line of sight. In that case, relativistic
beaming is not particularly effective, and the radio source should be
characterised by low variability and little or no other signatures of
an active nucleus as seen in the hosts of GPS galaxies, which appear
as passively evolving ellipticals (Snellen et al. 1998).



The reason that makes these radio galaxies small can be found in their
early evolutionary stage. From the ideas of Phillips and Mutel (1982),
who studied a number of ``Compact Doubles'', and  Carvalho (1985), one
can think of some unknown phenomenon that gives rise to the radio
activity in the nuclei of some giant elliptical galaxies. Relativistic
plasma is channeled into the radio jets, which start to dig their
way into the Inter Stellar Medium (ISM) of their host, first into the
Narrow Line Region (NLR) (corresponding to the GPS stage) and then
into the more tenuous and homogeneous ISM (the CSS stage) before
plunging into the Inter Galactic Medium (IGM) (extended radio source,
possibly with FR-II morphology), reaching projected linear sizes of
the order of hundreds of kpc or even more.  

The competing scenario of the ``frustration'' (van Breugel 1984), where
small radio sources are indeed old and remain confined within the host
galaxy by an ``anomalously'' dense interstellar medium, has been
excluded by the lack of any observational evidence of an ISM denser
than in the hosts of the extended radio sources (Fanti et al. 2000;
Gelderman 1996; de Vries et al. 1998; and many others) 

The observation of the increase in the separation of the outer edges
(hot-spots) in three Compact Symmetric Objects
(CSOs, i.e. the modern version of the ``Compact Doubles'') by Owsianik
\& Conway (1998) and Owsianik et al.(1998) definitely supported the
idea that these objects are young and growing.
Such motions have now been measured for a number of CSOs (see
Polatidis \& Conway, 2003), with projected speeds in the range
of 0.1-0.3~c. \\
Radiative ages determined on the basis of the integrated radio
spectrum are also consistent with the values derived from hot-spot
motions, just by assuming a simple model of continuous particle
injection and standard equipartition conditions (Murgia et al. 1999;
see also Murgia et al. 2003, for local spectral ageing
determination). 

The ``youth model'' has been outlined in some detail by various groups
(e.g. Fanti et al. 1995; Readhead et al. 1996; Begelman 1996; Snellen
et al. 2000); all agree on the most relevant aspects. The source
initially increases its luminosity (during the GPS stage) then, once
the radio emission leaves the NRL, expansion losses start to compete 
with synchrotron losses and the radio luminosity smoothly declines by
about one order of magnitude from the GPS to the extended radio
galaxy stage. \\ Synchrotron Self Absorption is thought to be
responsible for the turnover observed in the radio spectra. Bicknell
Dopita \& O'Dea (1997) proposed a model where the expanding lobe
creates a cocoon of shocked, ionised material that could account for
some amount of Free Free Absorption. Support for this possibility has
been found in a several cases (see Kameno et al. 2000 and 2003;
Mutoh et al. 2002).  

Typical examples of small and young radio sources are B2352+495 and
B0710+439, with a projected linear size of 160 and 120 pc
respectively, where the outer edges (hot-spots) have been measured to
separate at a speed of the order of 0.2~c, with a consequent age of
the order of a few in 10$^3$  yrs (Owsianik \& Conway 1998, see also
Polatidis \& Conway, 2003). On the other hand, another of these
small and young radio sources, namely B0108+388, is known to possess some
diffuse emission on the scale of tens of kpc. Although there is, as
yet, no definitive proof that this diffuse radio emission is indeed
physically related to the pc scale structure, it has been proposed
that the radio activity is a recurrent phenomenon and this extended
feature is the debris of a previous activity cycle. This
interpretation is also supported by the observation of recurrent
activity in giant radio galaxies (Schoenmakers et al. 2000).

By examining the diagram relating the intrinsic turnover frequency and
the projected linear size published by O'Dea (1998), it should then be
easy to find smaller and younger radio sources by searching for
convex spectra peaking at a few GHz or even higher frequencies. These
objects known as High Frequency Peakers (HFPs) are briefly discussed
in this paper.

\section{Rare gems ?}

It is easy to show that radio sources with spectra with a turn over at
very high frequencies (i.e. tens of GHz) are rare, since the initial
stages of radio emission are characterized by quite a rapid evolution
in terms of size and peak frequency. 

Let us consider a 10 yr-old source, sitting at z=1 and  growing at a
constant rate of 0.2~c, made up by two lobes and hot-spots; the jet
axis is in the plane of the sky. The source is then 0.61 pc (2.0 ly)
in size. To make the computation as simple as possible one 
can assume that the hot-spots are not very bright and that the bulk of
the radio emission comes from the two lobes, assumed to be homogeneous
ellipsoidal regions with a length of half the size of the source and
with an axial ratio of 0.25 (0.31 pc times 0.08 pc). One can assume a
total flux density of about 200 mJy at 22.5 GHz (the highest VLA
frequency in the HFP sample presented by Dallacasa et al. 2000)
equally distributed over two symmetric lobes. Assuming canonic
equipartition conditions, one would expect to observe the turnover
frequency at 38.4 GHz. Indeed this is a lower limit, given that using
a flux density in the optically thick region we underestimate the
total energy in the region and thus the turnover frequency.
It has also to be noted that radiative losses are very high given that
the relativistic electrons are located within very high magnetic
fields, and this would steepen considerably the optically thin
spectrum for sources having turnover frequencies below 22
GHz. Continuous injection of fresh realtivistic particles it is
expected to make this steepening less effective.

As the source ages the turnover moves down to 11.4 GHz for a 50
yr-old source (two lobes of about 1.5 pc in size) and then to 6.9 GHz
for a 100 yr-old source. The relative number of sources with peak flux
at about 40 GHz compared to those peaking at about 7 GHz will be 1 to
10. The turnover frequencies have also been computed for sources at
z=0.3, 0.5 and 2 (for z$>$~1.5-2 the turnover frequency does not
change significantly) and are reported in Table 1.

This simple calculation is complicated by the prediction of basically
all ``growth'' models that the radio source luminosity in the early
stage increases with size until the lobes advance further into the NRL,
before starting to decrease (the GPS becoming a CSS) proceeding even
further into the ISM and finally into the IGM.

Furthermore, the source components are known to be inhomogeneous: the
brightest regions are also the smallest and the turnover is somehow
higher than that calculated above. All this tends to make the convex
spectrum a bit broader than in the case of a pair of homogeneous
components; the spectral peak is mostly set by the region with the
larger contribution to the total flux density.

It should also be taken into consideration that for a given flux
density, the source at high redshift has an energy density and thus an
intrinsic peak frequency higher than a source located relatively
nearby, and therefore we are comparing objects with intrinsically
different luminosities (and magnetic field strengths also). 

In any case, if we neglect all the aforementioned effects, the message
coming from Table 1 is that generally the turnover moves at
frequencies below 10 GHz in a rather fast way if compared to the radio
source lifetime. 

An additional complication: in the very early days the equipartition
fields are very high (of the order of 1 Gauss or even stronger) and
so the relativistic particles are very shortlived. We thus expect
to observe only hot-spots and short backflow tails. This implies that
the emitting region is smaller than the size considered here to
estimate the physical parameters and so that the turnover frequency
may result somewhat higher than that reported in Table 1.

In summary, HFPs are rare and thus difficult to find; high frequency
catalogues (i.e. at cm wavelengths or even shorter) extended over
large sky areas are necessary to find newly born radio sources.  

\begin{table} 
\begin{center}
\begin{tabular}{c|rrr}
\hline
\hline
 redshift      &  ----- & $\nu_{turn}$ & ----- \\
       &          &  (GHz)  &         \\
       &  10yr    &  50yr   &    100yr\\
\hline
       &          &         &         \\
0.3  &    ~~24.6  &    7.3  &     4.4 \\
0.5  &    ~~30.3  &    9.1  &     5.4 \\
1.0  &    ~~38.4  &   11.4  &     6.9 \\
2.0  &    ~~42.9  &   12.9  &     7.7 \\
       &          &         &         \\
\hline
\hline
\end{tabular}
\end{center}
\caption{Observed turnover frequency observed for a source with a
  total flux density of 200 mJy at 22 GHz. The three columns on the
  right hand side refer to a source age of 10, 50 and 100 yrs
  respectively, corresponding to a linear size of 0.6, 3.0 and 6.0 pc}
\end{table}

\section{A brief history}
\begin{figure}
\begin{center}
\psfig{file=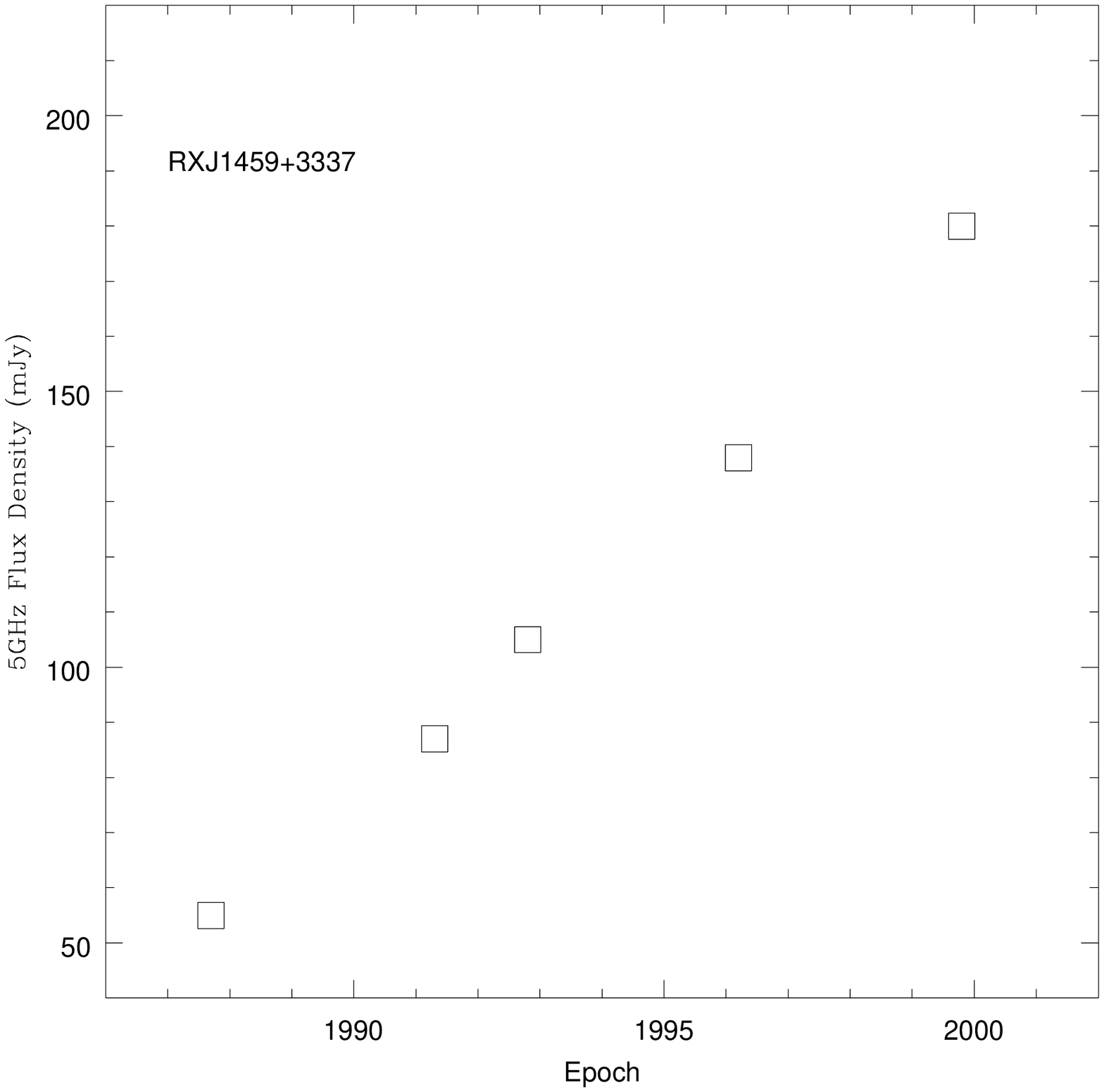, height=7.4cm}
\psfig{file=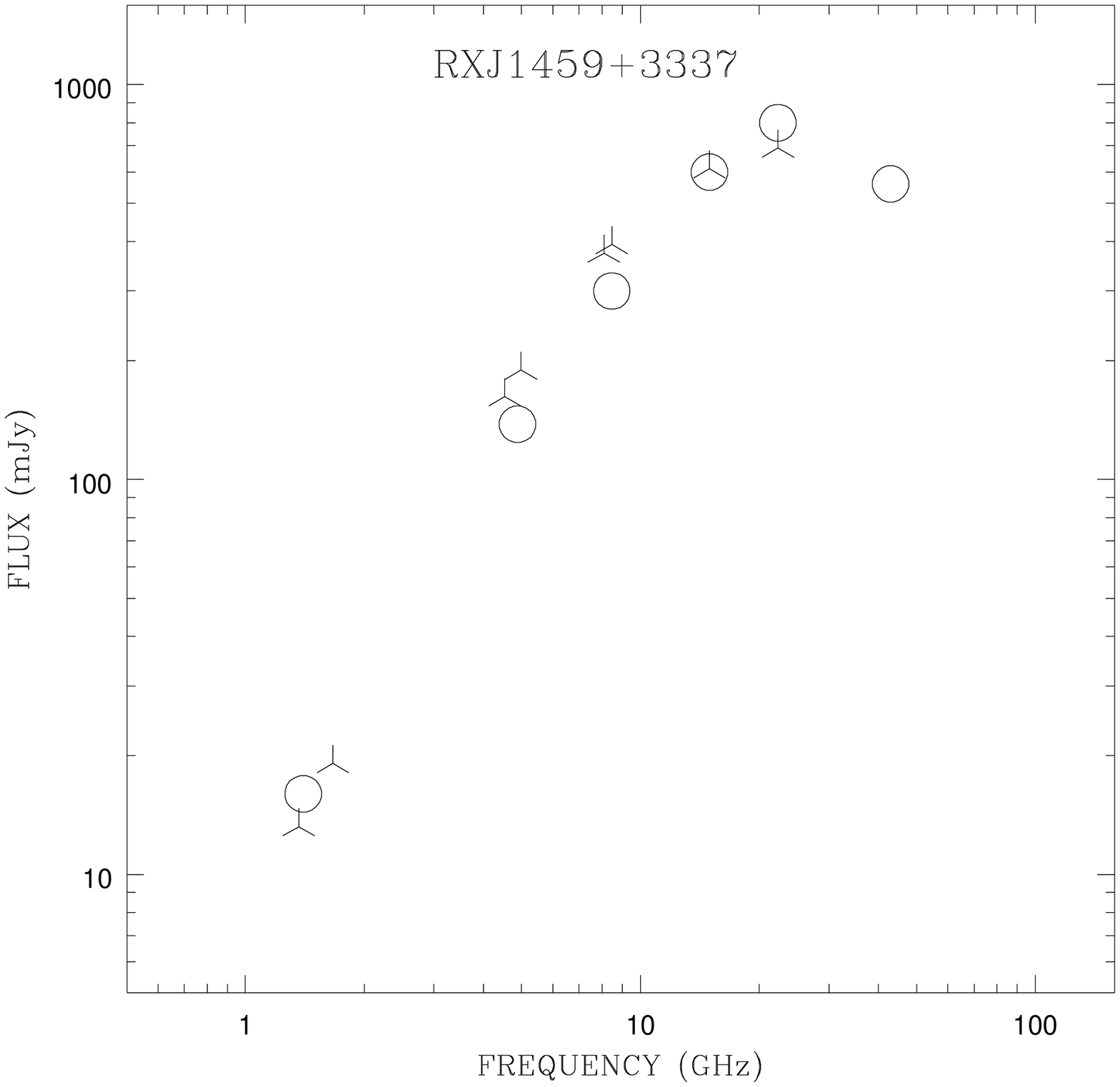, height=7.4cm}
\caption{RXJ1459.9+3337: ($left$) the total flux density at 5 GHz;
($right$) the V.LA radio spectrum in 1996 ({\it open circles}) and in
1999 ({\it triangular symbols}). Errors are generally smaller than the
size of the symbols.}
\label{RXJ}            
\end{center}
\end{figure}

Edge et al.(1996) noted the existence of GPS sources with a
turnover frequency higher than usually found in the samples of
Stanghellini et al. (1990, 1998) and Spoelstra, Patnaik \&
Gopal-Krishna (1985). In particular RXJ1459.9+3337 (a quasar at
z=0.65) had a number of observations at 5 GHz showing a steady
increase in the total flux density over a period of about 10 yrs,
starting at around 50 mJy at the end of 1987 and reaching about 140 mJy
in 1996. The ``simultaneous'' radio spectrum from the VLA showed a
radio spectrum with a peak around 30 GHz (about 50 GHz in the rest
frame). The source has been observed again with the VLA in 1999 since
the source belongs to the ``faint'' HFP sample (Stanghellini et al.
in preparation; the ``faint'' sample covers the flux density interval
between 50 and 300 mJy in the GB6 catalogue), and the observed flux
density at 5 GHz sits exactly on the extrapolation from Edge et
al.'s plot (see Fig. \ref{RXJ}~$left$). It is also interesting to note
that the spectral peak is likely to have moved down to a slightly
lower frequency, possibly indicating that the source and/or component
expansion is decreasing the optical depth below the turnover
frequency. However, the 1999 VLA data lack the 43 GHz measurements,
and thus it is not possible to derive a firm estimate for the spectral
peak at this epoch.

The possibility that the steady increase of flux density with time is
only apparent has also to be taken into account, and could be due to
the sparse time sampling of a more irregular flux density variability
with the typical ``random'' pattern, that, however, appears to be
linear due to the sampling carried out with a certain amount of
``cosmic conspiracy''. 

\section{A complete sample of HFPs }

As mentioned above the ideal, intrinsically small and young radio
sources are those identified with galaxies. However, it is nearly
impossible to select samples by starting from optical identification,
particularly when rare objects are searched for. Then the conventional
approach to select samples on the basis of the radio spectrum (the
usual convex, bell-shape with a clear optically thin region) have
to be used. 

Among the signatures of young radio sources, the turnover frequency
above a few GHz is probably the most characterstic of this class, and
also the easiest to investigate.
Then, by comparing the NVSS (Condon et al. 1998) with the GB6
(Gregory et al. 1996) catalogues one would expect that the young
sources must have a rather inverted spectrum and must also appear
unresolved on both surveys. However, this fairly simple selection is
spoiled by the relatively long time lag between the GB6 and the VLA
observations. In fact, variable sources at a high state of activity
during the GB6 campaign, are selected in this way as well. Therefore
it was definitely necessary to carry out simultaneous multifrequency
observations to define the spectral shape of all the candidates. 

Dallacasa et al. (2000) defined a sample of ``bright'' HFPs, with flux
density in excess of 300 mJy in the GB6 catalogue and inverted
spectra (steeper than 0.5 if S$\propto \nu^{-\alpha}$) between 1.4
(the NVSS measurement) and 4.9 GHz. 
About 100 candidates were then observed at the VLA to get
simultaneous radio spectra between 1.4 and 22 GHz and then separate
variable flat spectrum sources from those with intrinsically
convex-shaped radio spectra. The final sample of genuine HFP sources
consists of 55 objects and covers the area of the GB6 catalogue,
although a few small areas have not yet been covered by the NVSS. 

About 1750 sources brighter thant 300 mJy in the GB6 were found in the
NVSS and the final number of HFPs represents a rather conspicuous
fraction (3\%) if compared to the expectations, bearing in mind a
model of a young radio source quite rapidly developing within the host
galaxy.  The selection does not impose any condition on the optical
identification. Indeed we should restrict our attention to galaxies
only, possibly also rejecting those broad line hosts classified as
galaxies given that they are indeed relatively nearby and it is
possible to reveal consistent starlight around them.
About half of the bright HFPs already had been identified in the
literature and a project to complete such information is continuing
(Dallacasa, Falomo \& Stanghellini 2002). Most of the HFPs are
identified with high redshift quasars and only about 25\% (15 objects
in total, including two nearby broad line radio galaxies) can be
considered non quasar or BL Lac. 

Therefore, the intrinsically young radio sources (i.e. narrow line
galaxies) become less than 1\% and it is no surprise to find that the
intrinsic peak frequency is generally much smaller than in HFP quasars
also as a consequence of the different redshift range spanned by the
two populations, with the majority of the quasars found at redshift
larger than 1.5.

The expected number of very young (about 100 yr old) sources can be
calculated from the ratio of their typical age and the average age of
extended radio sources (about 10$^7$ yr). This would imply a fraction
of the order of 0.01\%, but we must also consider that in all the
``youth'' models the sources progressively decrease their radio
luminosity as the mini lobes leave the NLR, and then we should have
considered a larger sample of ``old'' radio sources by decreasing the
flux density limit by about one order of magnitude. In this case,
nearly all the 54000 sources in the GB6 would form the ``old'' radio
source sample to be used in the determination of the fraction of young
radio sources in catalogues.

Among the 31 sources with measured redshift only 11 have an intrinsic
peak frequency above 22 GHz, and all of them are quasars. There are
two sources more with observed peak frequency beyond 22 GHz, and both
are identified with stellar objects (redshift is not yet available).
The few galaxies with measured or estimated redshift have peak
frequencies in the range 5 -- 12 GHz.

The radio morphology is also an important tool to assess
whether a source is indeed young (lobe dominated) or just a knot in a
jet which is dominant over the entire radio emission (as in 4C39.25,
Alberdi et al. 2000), and in fact the CSO or MSO (Medium-size
Symmetric Object) samples also contain a few QSOs. VLBA imaging in the
optically thin part of the spectrum is a useful tool in this respect
(see also Tinti et al. 2003). 

\subsection{Polarisation} 
\begin{figure}
\begin{center}
\psfig{file=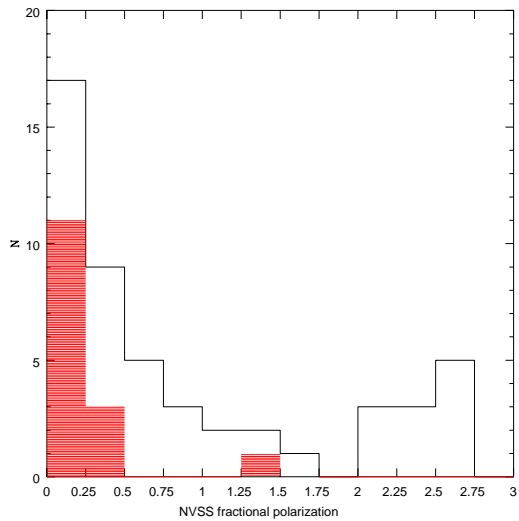, height=7.4cm}
\caption{NVSS fractional polarisation: galaxies are represented by
shaded areas, quasars by the unshaded areas; most of the values below 0.5\%
must be considered as upper limits; the 2.5--2.75\% bin contains
all the sources with fractional polarisation in excess of 2.5\%}
\label{POL}            
\end{center}
\end{figure}
Intrinsically small sources are in regions close to the active nucleus
where a substantial amount of high energy radiation is produced and
where ionisation and magnetic field strength are important.
Consequently the Faraday Rotation is likely to be significant and
where the Faraday screen is not homogeneous and not resolved by the
observations, then the (beam) depolarisation becomes dominant at
relatively high frequencies.  

Among the bright HFPs, the polarization information is still not
available in a systematic way except in the NVSS, which happens to be
in the optically thick part of the spectrum of all the
sources. However, it is possible to distinguish the behaviour of
the HFP quasars (and the two BL Lacs) from the other HFPs identified
with galaxies or still empty fields. As can be be seen in
Fig. \ref{POL}, the former span a wide range of fractional polarisation,
with measured values up to 5\%, while the latter are not polarised at
all with one exception.  This is what it is expected in case that HFP
quasars are mostly larger radio sources seen with the jet aligned to
the line of sight and thus the radiation comes from outside the region
with high local ionisation. In a study concerning CSS and GPS sources
done in a similar way Cotton et al. (2003) find that there is a
quite sharp change in the fractional polarisation which rises to
significant values once the source emerges from a region of about 2-3
kpc in radius. Given that HFPs are well within this region, they are
expected to be unpolarised. 

\subsection{Variability}
It is well known that  flux density variability may substantially
modify the observed spectrum of blazars in all the bands of the
electromagnetic spectrum, sometimes even challenging the optical
identification (e.g. BL Lac itself, Vermeulen et al. 1995).

Little is known about the radio variability in HFP sources, except for
a very few objects already known for other reasons: the HFP galaxy
B0108+388 (also in the Stanghellini et al. 1998 GPS sample) is known
to possess a rather constant flux density at all radio frequencies;
the  BL Lac B1749+096 and the quasars B0923+392 (4C39.25) and
B2134+004 (the latter is also in the Stanghellini et al. 1998 GPS
sample) are known to possess some amount of flux density and
polarisation variability (see the VLA/VLBA polarisation calibration
webpage). On the other hand, the broad line galaxy B1404+286 (OQ208,
also a GPS in the Stanghellini et al. 1998 sample) showed a steady
decrease of its flux density at 5 GHz in the eighties, but now this
decrease has stopped. The flux density at 1.4 GHz, however, did not
show any significant variability. 

\begin{figure}
\begin{center}
\psfig{file=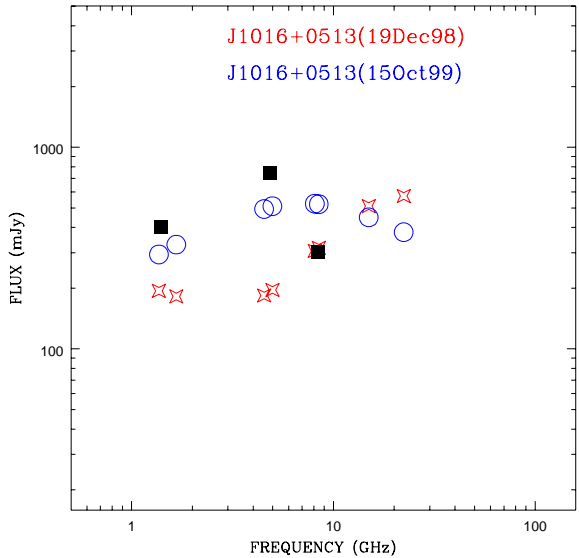, height=8.0cm}
\caption{Flux density and spectral variability in the HFP source
  J1016+0513. VLA data are represented in stars and circles, while
  NVSS, 87GB and JVAS flux densities are represented by the filled
  squares} 
\label{VAR}            
\end{center}
\end{figure}

For most of the 55 HFP sources in the bright sample there is only the
simultaneous multifrequency VLA datapoints in Dallacasa et al. (2000),
the NVSS, the GB6 and the JVAS measure at 8.4 GHz, where it is however
possible to see a significant amount of flux density variability
(Dallacasa et al. 2000).
The VLA multifrequency observations of HFP candidates were split into
several runs separated by weeks or months, and a handful of sources
were observed twice. Among these sources, J1016+0518 (B1013+054)
shows a remarkable change: the two simultaneous radio spectra are
shown in Fig. \ref{VAR}.
At both epochs, the overall radio spectrum maintains a convex shape,
although the first epoch is characterised by a complex structure. The
peak has moved from $>$22 GHz down to 7 GHz over about 10 months. If
we assume classical equipartition conditions and use O'Dea's (1998)
plot, we can estimate the size of the source by means of the 
spectral peak. The increase in size is then evaluated to be from 1.0
to 3.0 pc and this would correspond to an expansion velocity of about
16~c, which is clearly unrealistic, i.e. we cannot interpret this
source in terms of a young object where the lobes are expanding within
the host galaxy, but rather as a beamed object whose radio emission is
dominated by a single knot in the jet.

In this respect HFP quasars can be considered the counterpart of
steep-spectrum radio quasars (SSRQ) in the population of powerful and
extended radio sources, i.e. objects oriented at angles intermediate 
to the line of sight and where relativistic beaming is effective
although not so strongly as in blazars.

Further VLA polarisation-sensitive multifrequency observations are
under examination in order to evaluate the incidence of variability
(also in spectral shape) among HFP sources and also to estimate the
degree of polarisation in the optically thin part of the spectrum. If
significant polarisation is also detected below the turnover
frequency, these data will allow one to study the mechanism
responsible for self-absorption (see also Kameno et al. 2003). 

\section{Summary}

The hosts of the ``bright'' HFPs are mostly quasars (with two BL Lac
objects), and the galaxies are rather rare, as expected from the
estimate of the radiative age as derived from the spectral peak.
Only 15 out of the 55 bright HFP sources can be considered galaxies,
although two are classified as Broad Line galaxies, and the final
number of prototype young radio sources is likely to be smaller.
Accurate VLBI morphological studies in the optically thin part of the
spectrum play a key role in the correct classification.

If we consider Unified Scheme models for powerful (type 1) but
{\bf small} radio loud objects, we could think of a sequence of
sources with the radio jets oriented at decreasing angles to the line
of sight formed by HFP/GPS/CSS galaxies, then HFP/GPS/CSS quasars and
finally a small number of OVV/FSRQ/BLAZARS.

Indeed there could be some bias towards HFP/GPS/CSS quasars: these are
likely to be dominated by a knot in the jet were some degree of
relativistic beaming increases the apparent luminosity of the object,
and therefore high frequency catalogues are biased towards them and
against galaxies.

%
%





\section*{Acknowledgments}
I am indebted to Derek Jones for careful reading the manuscript and
substantially improving the English. \\
I acknowledge financial support by the Italian MIUR under grants
COFIN-2001-02-8773 and COFIN-2002-8118\\
The National Radio Astronomy Observatory (NRAO) is a facility of the
National Science Foundation operated under a cooperative agreement by
Associated Universities, Inc.\\
This research has made use of the NASA/IPAC Extragalactic Database
(NED), which is operated by the Jet Propulsion Laboratory, Caltech,
under contract with the National Aeronautics and Space Administration.

\section*{References}





\reference Alberdi, A., G\'omez, J. L., Marcaide, J. M., Marscher,
A. P.\and P\'erez-Torres, M. A., 2000, A\&A, 361, 529
\reference Begelman, M.C. 1996, in Proc. Cygnus A - Study of radio
Galaxy, Carilli C.L. \and Harris D.E. (eds.) CUP, Cambridge, p. 209 
\reference Bicknell, G. V., Dopita, M. A.,\and O'Dea, C. P. 1997, ApJ,
485, 112
\reference van Breugel, W. J. M. 1984, in Proc. IAU Symp. 110, Fanti,
R., Kellerman, K.I. \and Setti, G., (eds), Reidel, Dordrecht, 59
\reference Carvalho, J. C. 1985, MNRAS, 215, 463
\reference Condon, J.J., Cotton, W.D., Greisen, E.W., et al., 1998, AJ
115, 1693  
\reference Cotton, W.D., Dallacasa, D., Fanti, C., Fanti, R., Foley,
A.R., Schilizzi, R.T., Spencer, R., Saikia, D.J., Garrington, S. 2003,
PASA, 
\reference Dallacasa, D., Falomo, R.\and Sanghellini, C. 2002, 
A\&A, 382, 53
\reference Dallacasa, D., Stanghellini, C., Centonza, M. \and Fanti,
R. 2000, A\&A, 363, 887
\reference Edge, A.C., Jones, M., Saunders, R., Pooley, G. \and Grainge,
K. 1996, in Proc. 2$^{nd}$ workshop on GPS and CSS radio sources, 20, 12
Snellen,I.A.G. et al. (eds.), Leiden, 43
\reference Fanti, C., Fanti, R., Dallacasa, D., Schilizzi, R. T.,
Spencer, R. E. \and Stanghellini, C. 1995, A\&A, 302, 317 
\reference Fanti, C., Pozzi, F., Fanti, R., Baum, S. A., O'Dea, C. P.,
et al. 2000, A\&A, 358, 499
\reference Gelderman, R. 1996, in Proc. 2$^{nd}$ workshop on GPS and
CSS radio sources, Snellen,I.A.G. et al. (eds.), Leiden
\reference Gregory, P.C., Scott, W.K., Douglas, K. \and Condon, J.J., 1996,
ApJS 103, 427 
\reference Kameno, S., Horiuchi, S., Shen, Z. Q., Inoue, M.,
 Kobayashi, H., Hirabayashi, H. \and Murata, Y. 2000, PASJ, 52, 209
\reference Kameno, S., et al. 2003, PASA, 20, in press
\reference Murgia, M., Fanti, C., Fanti, R., Gregorini, L., Klein, U.,
Mack, K.-H. \and Vigotti, M. 1999, A\&A, 345, 769
\reference Murgia, M. 2003, PASA, 20, 19
\reference Mutoh, M., Inoue, M., Kameno, S., Asada, K., Kenta, F. \and
Uchida, Y. 2002, PASJ, 54, 131
\reference O'Dea, C. P. 1998, PASP 110, 493
\reference Owsianik, I., Conway, J.E. \and Polatidis, A. G. 1998,
A\&A, 336, L37  
\reference Owsianik, I. \and Conway, J.E. 1998, A\&A, 337, 69 
\reference Polatidis, A.G., Conway, J.E. 2003, PASA, 20, in press
\reference Phillips, R.B., \and Mutel, R. L. 1982, A\&A, 106, 21
\reference Readhead, A.C.S., Taylor, G.B., Pearson, T.J. \and  Wilkinson,
P.N. 1996, ApJ 460, 634  
\reference Schoenmakers, A. P., Mack, K.-H., de Bruyn, A. G.,
R\"ottgering, H. J. A., Klein, U. \and van der Laan, H. 2000, MNRAS, 315,
371 
\reference Snellen, I. A. G., Bremer,  Schilizzi, R. T.,  Miley,
G. K. \and van Ojik, R. 1998, MNRAS, 279, 1294
\reference Snellen, I. A. G., Schilizzi, R. T.,  Miley, G. K., de
Bruyn, A. G.,  Bremer, M. N. \and R\"ottgering, H. J. A 2000, MNRAS, 319, 445
\reference Spoelstra, T. A. T., Patnaik, A. R. \and Gopal-Krishna 1985,
A\&A, 152, 38 
\reference Stanghellini, C., Baum, S. A., O'Dea, C. P. \and  Morris,
G. B. 1990, A\&A 233, 379 
\reference Stanghellini, C., O'Dea, C. P.,  Dallacasa, D., Baum,
 S. A., Fanti, R. \and Fanti, C. 1998 A\&AS 131, 303
\reference Tinti, S., Dallacasa, D., Stanghellini, C., Celotti,
A. 2003, PASA, 20, in press
\reference Urry, C. M. \and Padovani, P. 1995, PASP, 107, 803 
\reference Vermeulen, R. C., Ogle, P. M., Tran, H. D., Browne, I. W. A.,
Cohen, M. H., Readhead, A. C. S., Taylor, G. B. \and Goodrich,
R. W. 1995, ApJL, 452, 5
\reference de Vries, W. H., O'Dea, C.P., Perlman, E., Baum, S.A., et
al. 1998, ApJ, 503, 138


\end{document}